\documentclass[pra,twocolumn,groupedaddress,superscriptaddress,showpacs,amsmath,amssymb,a4paper]{revtex4}
\usepackage{geometry}
\usepackage{graphicx}
\usepackage{mathrsfs}
\usepackage{amssymb}
\usepackage{amsmath}
\usepackage{amsthm}
\usepackage{amsbsy}
\usepackage{bm}
\usepackage{hyperref}

\theoremstyle{definition}

\newtheorem{example}{Example}

\newcommand{\bra}[1]{\langle #1|}
\newcommand{\ket}[1]{| #1 \rangle }
\newcommand{\ip}[2]{{\langle #1|}{ #2 \rangle }}

\newcommand{\be}{\begin{eqnarray}}
\newcommand{\ee}{\end{eqnarray}}

\newcommand{\blockmatrix}[3]{%These end of the line comments are neccessary
\begin{minipage}[t][#2][c]{#1}%
\center%
#3%
\end{minipage}%
}%

\begin{document}

\title{Quantum evolution in the stroboscopic limit of repeated measurements}

\author{I. A. Luchnikov}

\affiliation{Moscow Institute of Physics and Technology,
Institutskii Per. 9, Dolgoprudny, Moscow Region 141700, Russia}

\author{S. N. Filippov}

\affiliation{Moscow Institute of Physics and Technology,
Institutskii Per. 9, Dolgoprudny, Moscow Region 141700, Russia}

\affiliation{Institute of Physics and Technology, Russian Academy
of Sciences, Nakhimovskii Pr. 34, Moscow 117218, Russia}

\begin{abstract}
We consider a quantum system dynamics caused by successive
selective and non-selective measurements of the probe coupled to
the system. For the finite measurement rate $\tau^{-1}$ and the
system-probe interaction strength $\gamma$ we derive analytical
evolution equations in the stroboscopic limit $\tau \rightarrow 0$
and $\gamma^2 \tau = {\rm const}$, which can be considered as a
deviation from the Zeno subspace dynamics on a longer timescale $T
\sim (\gamma^2 \tau)^{-1} \gg \gamma^{-1}$. Non-linear quantum
dynamics is analyzed for selective stroboscopic projective
measurements of an arbitrary rank. Non-selective measurements are
shown to induce the semigroup dynamics of the system-probe
aggregate. Both non-linear and decoherent effects become
significant at the timescale $T \sim (\gamma^2 \tau)^{-1}$, which
is illustrated by a number of examples.
\end{abstract}


\maketitle

%----------------------------------------------------------------------

\section{\label{section-introduction} Introduction}

Measurements naturally provide some information about the system
involved. In quantum physics, no information can be gained without
disturbance of the system~\cite{heinosaari-ziman,clerk-2010}. The
ultimate form of such a disturbance takes place in projective
measurements, when the observation of a particular outcome $i$
leads the system into the state $\ket{\psi_i}$ (conditional state
preparation). The mathematical form of the noise and disturbance
relation has been recently found for general fuzzy
observables~\cite{buscemi-2014}. The effect of system disturbance
becomes more visible in sequential and repeated measurements.

By repeated measurements we mean successive measurements of the
same quantum observable. For instance, a high repetition rate of
the same projective measurement results in the quantum Zeno
effect~\cite{misra-sudarshan-1977}, in which the system state
dynamics is frozen. Non-projective repeated measurements at finite
frequency may also lead to a perfect freezing~\cite{layden-2015}.
However, the accelerated decay is more ubiquitous in the case of a
slow repetition rate of measurements (anti-Zeno
effect)~\cite{kofman-2000,fischer-2001,facchi-2001}. Since the
conditional output state is non-linearly related with the input
density operator, other non-trivial dynamics is possible including
the emerging chaotic behavior~\cite{kiss-2006,kiss-2011}.
Repetitive measurements enable maintenance of quantum coherence in
the presence of noise~\cite{konrad-2012} or acceleration of
decoherence~\cite{facchi-2004}. Repeated selective measurements
are applicable in ground state cooling~\cite{li-2011}. Repetition
of non-selective measurements at particular time moments allows
controlling the probability of transitions between qubit
levels~\cite{pechen-2015}.

By sequential measurements we mean successive measurements of
different non-commuting quantum observables. Statistics of general
sequential measurements may exhibit the properties of
undecidability~\cite{eisert-2012}, universality with respect to
the construction of a joint observable~\cite{heinosaari-2015}, and
informational completeness for state
tomography~\cite{lorenzo-2013,diosi-2016}. Moreover, sequential
measurements find applications in estimation of quantum system
parameters~\cite{burgarth-2015}, process
tomography~\cite{bassa-2015}, one-way quantum
computing~\cite{raussendorf-2001}, channel
decoding~\cite{lloyd-2011}, and detection~\cite{suzuki-2016} and
generation~\cite{coles-2014} of non-classical correlations.

Periodic interventions in the quantum system evolution via unsharp
measurements are analyzed in a series of
papers~\cite{audretsch-1997,korotkov-2001,audretsch-2001,audretsch-2002,ai-2013}.
In Ref.~\cite{audretsch-1997}, restricted path integrals are used
in the phenomenological treatment, which is equivalent to the
introduction of non-Hermitian Hamiltonians. A particular physical
realization of continuous measurements is considered in
Ref.~\cite{korotkov-2001}. In
Refs.~\cite{audretsch-2001,audretsch-2002}, two-outcome
measurements of a qubit system are considered and the difference
equations on the density operator are obtained by approximating
the binomial distribution of outcomes by the Gaussian form. In
Ref.~\cite{ai-2013}, indirect measurements of a decaying system
are realized via auxiliary states, with such coupling-based
measurements increasing the decay rate.

In present paper, we develop the ideas of indirect measurements by
introducing a probe, which interacts with the system and is
periodically measured in equal time intervals $\tau$. The primary
goal is to analyze the system dynamics caused by repeated
measurements. The system-probe interaction Hamiltonian is
arbitrary, with $\gamma$ being the characteristic coupling
strength. Our aim is to study the measurement-induced system
dynamics at a timescale $T \sim (\gamma^2\tau)^{-1}$ in contrast
to the interaction timescale $T_{\rm int} \sim \gamma^{-1}$. We
show that in the \emph{stroboscopic limit} $\tau \rightarrow 0$,
$\gamma^2 \tau = {\rm const}$, the resulting evolution allows an
analytical solution for both selective and non-selective
measurements. The stroboscopic limit implies $\gamma \tau \ll 1$,
so the derived analytical solutions are valid at time $T \gg
T_{\rm int}$ and can be interpreted as deviations from the Zeno
dynamics at longer times. Note that we consider stroboscopic
dynamics, which differs from so-called stroboscopic
non-demolishing measurements of periodic
quantities~\cite{caves-1980,braginsky-1992,jordan-2005,ruskov-2005}.
Conceptually, our approach is similar to that in
Ref.~\cite{layden-2016}, where the coupling dynamics is intervened
by resets of the probe state or by renewal of the environment (cf.
the collision model~\cite{ziman-2011,rybar-2012}). However, the
non-selective measurements cannot be reduced to resets, which
differs in our model from those studied previously. Interestingly,
the stroboscopic condition $\gamma^2 \tau = {\rm const}$ resembles
the analogous condition in the stochastic limit for calculating
dominating contributions of the open system dynamics at long
times~\cite{accardi-2002}.

The paper is organized as follows. In Sec.~\ref{section-problem},
we formulate the problem of quantum evolution due to indirect
stroboscopic measurements. In Sec.~\ref{section-rank-1}, we show
that frequent observations of the probe via selective rank-1
projective measurements ($\gamma \tau \ll 1$) effectively freeze
the probe evolution (Zeno subspace effect~\cite{facchi-2002}) but
the system evolution is non-linear and can be described by the
analytical effective Hamiltonian at a timescale $T \sim
(\gamma^2\tau)^{-1}$. The introduced stroboscopic limit differs
from the conventional limit in Zeno subspace
effect~\cite{facchi-2002,li-2013} and describes longer times when
non-unitary effects become significant. In
Sec.~\ref{section-rank-r}, projective measurements of an arbitrary
rank $r$ are considered. Such measurements enable the probe to
evolve non-trivially within the measurement-invariant subspace,
which affects the system evolution too. In
Sec.~\ref{section-non-selective}, general non-selective
measurements of the probe are considered and a semigroup property
of the system-probe dynamics is derived. In
Sec.~\ref{section-conclusions}, brief conclusions are outlined.

%----------------------------------------------------------------------

\section{\label{section-problem} Dynamics under stroboscopic selective measurements}

Let ${\cal H}_{\rm sys}$ and ${\cal H}_{\rm pr}$ be the system and
probe Hilbert spaces, respectively. For the sake of simplicity we
assume that both ${\cal H}_{\rm sys}$ and ${\cal H}_{\rm pr}$ are
finite dimensional. By ${\cal B}({\cal H}_{\rm sys})$ and ${\cal
B}({\cal H}_{\rm pr})$ denote the linear spaces of operators
acting on ${\cal H}_{\rm sys}$ and ${\cal H}_{\rm pr}$,
respectively. Any system-probe Hamiltonian $H$ admits the
resolution
\begin{equation}
H=\gamma\sum_{j}{A_j\otimes B_j},
\end{equation}

\noindent where the dimensionless operators $A_j \in {\cal
B}({\cal H}_{\rm sys})$ and $B_j \in {\cal B}({\cal H}_{\rm
probe})$ have operator norms $\| A_j \|_{\infty}, \| B_j
\|_{\infty} \leqslant 1$, with $\| A \|_{\infty} = \max_{\psi: \,
\ip{\psi}{\psi} = 1} \sqrt{ \bra{\psi} A^{\dag} A \ket{\psi} }$.
The parameter $\gamma$ defines a characteristic strength of the
system-probe interaction. Hereafter we assume the Planck constant
$\hbar = 1$, so energy has the dimension of frequency.

Let $\varrho$ be the aggregate density operator on ${\cal H}_{\rm
sys} \otimes {\cal H}_{\rm pr}$, which describes the system and
probe altogether ($\varrho^{\dag} = \varrho \geqslant 0$, ${\rm
tr}[\varrho] = 1$). Then the unitary evolution of duration $t$
reads
\begin{equation}
\label{unitary} {\cal U}_t [\varrho] = e^{-iHt} \, \varrho \,
e^{iHt}.
\end{equation}

\begin{figure}
\includegraphics[width=8.5cm]{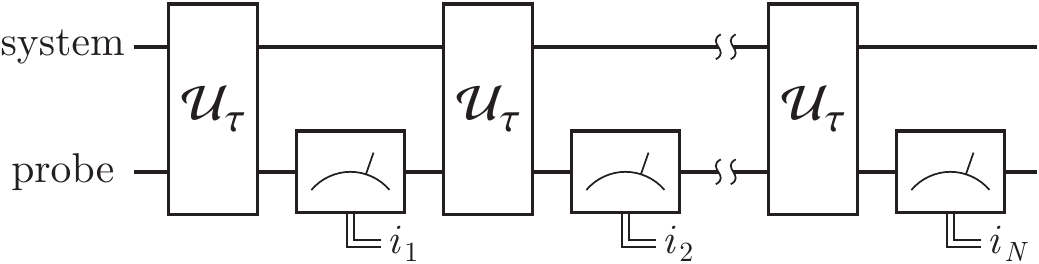}
\caption{\label{figure1} Stroboscopic selective measurements of
the probe result in a non-linear dynamics of the system.}
\end{figure}

\begin{figure}
\includegraphics[width=6cm]{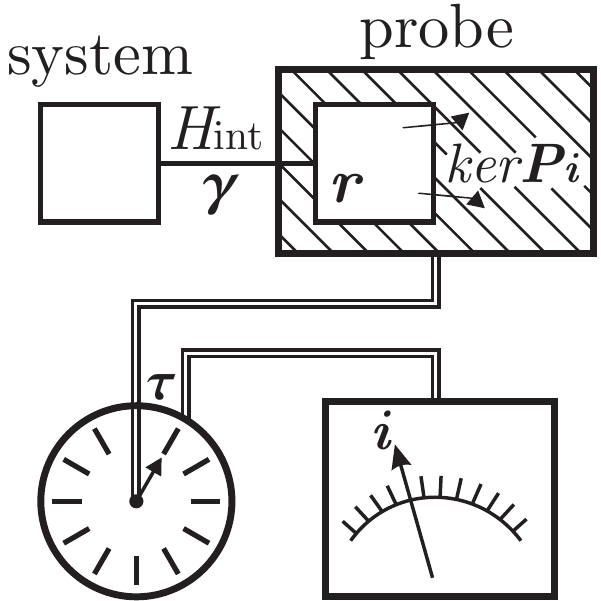}
\caption{\label{figure2} Schematic of stroboscopic evolution.
Probe is being measured at successive moments with equal time
intervals $\tau$ between them. Characteristic strength of the
system-probe interaction is $\gamma$. Rank-$r$ measurements of the
probe restrict the system-probe evolution to the subspace ${\cal
H}_{\rm sys} \otimes {\cal H}_r$, where ${\cal H}_{\rm pr} = {\cal
H}_r \oplus {\rm ker} P_i$.}
\end{figure}

The probe is being measured repeatedly after equal time intervals
$\tau$, see Figs.~\ref{figure1} and \ref{figure2}. We will refer
to such an interrupted dynamics as a \emph{stroboscopic}
evolution. Note that this concept differs from stroboscopic
measurements discussed
in~\cite{caves-1980,braginsky-1992,jordan-2005,ruskov-2005}. By
time $t=T$ the number of performed measurements equals $N =
\lfloor T/\tau \rfloor$. If all those measurements resulted in the
outcomes $i_1, \ldots, i_N$ sequentially, then the system-probe
transformation is described by the following trace-decreasing map:
\begin{equation}
\label{instrument-exact-arbitrary} \Phi_{T} = {\cal U}_{T - N
\tau} \circ {\cal I}_{i_N} \circ \ldots \circ {\cal U}_{\tau}
\circ {\cal I}_{i_2} \circ {\cal U}_{\tau} \circ {\cal I}_{i_1},
\end{equation}

\noindent where $\circ$ denotes concatenation of maps and ${\cal
I}_{i_n}$ is the instrument describing the system-probe
transformation if the outcome $i_n$ is observed. Namely, the
instrument is a completely positive trace-decreasing map such that
the conditional output state is
\begin{equation}
\label{conditional-output-state} \varrho_i = \frac{{\cal I}_i
[\varrho] }{ {\rm tr}\big[ {\cal I}_i [\varrho] \big]}.
\end{equation}

Probability of realization of the map
\eqref{instrument-exact-arbitrary} for a given initial density
operator $\varrho$ equals $p_{\rm \Phi}={\rm tr}\big[\Phi_T
[\varrho]\big]$, where ${\rm tr}[\varrho]=1$.

Tracing out the probe, we get the system density operator
evolution
\begin{equation}
\varrho_{\rm sys} (T) = {\rm tr}_{\rm pr} \left\{ \frac{\Phi_{T} [
\varrho(0) ]}{{\rm tr} \big[\Phi_{T} [ \varrho(0)] \big]}
\right\}.
\end{equation}

%%%%%%%%%%%%%%%%%%%%%%%%%%%%%%%%%%%%%%%%%%%%%%%%%%%%%%%%%%%%%%%%%%%
%%%%%%%%%%%%%%%%%%%%%%%%%%%%%%%%%%%%%%%%%%%%%%%%%%%%%%%%%%%%%%%%%%%

We consider in detail the situation when the outcomes $i_1,
\ldots, i_N = i$ are all coincident. Physically, the probability
of such a sequence of outcomes is quite high if $\gamma\tau \ll 1$
because this case corresponds to the quantum Zeno effect for the
probe. This scheme resembles the repetition of pre- and
post-selected measurements with identical pre- and post-selected
states of the probe (state $\ket{i}$, see, e.g., the
review~\cite{kofman-2012}), however, no actual measurement of the
system state is performed. Instead, the induced system dynamics is
the primary goal of the study.

Observation of coincident outcomes $i_k = i$, $k=1, \ldots, N$
results in the following transformation:
\begin{equation}
\label{instrument-exact} \Phi_{T} = {\cal U}_{T - \lfloor T/\tau
\rfloor \tau} \circ \left( {\cal U}_{\tau} \circ {\cal I}_i
\right)^{\lfloor T/\tau \rfloor}.
\end{equation}

For projective measurements the instrument ${\cal I}_i$ takes the
form of a trace-decreasing map with a single Kraus operator:
\begin{equation}
\label{C-P} {\cal I}_i [\varrho] = C_i \varrho C_i, \qquad C_i =
I_{\rm sys} \otimes P_i,
\end{equation}

\noindent where $P_i$ is a projector $P_i = P_i^2 \in {\cal
B}({\cal H}_{\rm pr})$. We will refer to $r={\rm rank}P_i$ as the
rank of the measurement. If $r=1$, then $P_i =
\ket{\varphi}\bra{\varphi}$ and the observation of outcome $i$
means that the probe state reduces to
$\ket{\varphi}\bra{\varphi}$. If $r={\rm dim}{\cal H}_{\rm pr}$,
then such a measurement is completely uninformative as $P_i =
I_{\rm pr}$, the identity operator on the probe Hilbert space (no
measurement in fact). The intermediate case $1 < r < {\rm
dim}{\cal H}_{\rm pr}$ leaves some freedom for the probe evolution
in the subspace ${\cal H}_r = {\rm supp} P_i$, where ${\rm supp}
P_i$ denotes the support of operator $P_i$ (see
Fig.~\ref{figure2}).

It is natural to suppose that the initial state of the system and
probe is factorized, i.e. $\varrho(0) = \varrho_{\rm sys}(0)
\otimes \varrho_{\rm pr}(0)$, with ${\rm supp}\varrho_{\rm pr}(0)
\subseteq {\rm supp} P_i$, which is guaranteed by the first
measurement. The system evolution
\begin{equation}
\label{system-exact-evolution} \varrho_{\rm sys} (t) = {\rm
tr}_{\rm pr} \left\{ \frac{\Phi_{t} [ \varrho_{\rm sys}(0) \otimes
\varrho_{\rm pr}(0) ] }{ {\rm tr} \big[ \Phi_{t} [ \varrho_{\rm
sys}(0) \otimes \varrho_{\rm pr}(0) ] \big]} \right\}
\end{equation}

\noindent is non-linear if $H$ and $C_i$ do not commute. Even
though Eq.~\eqref{system-exact-evolution} is rather complicated,
the approximate analytical solution can be found if the coupling
strength $\gamma$ and the stroboscopic period $\tau$ are
appropriately related. In what follows, analytical solutions are
derived and compared with the exact ones for various ranks $r$ of
the projector $P_i$.

%----------------------------------------------------------------------

\subsection{\label{section-rank-1} Analytical solution for rank-1 projectors}

If $r=1$, then $C_i = I_{\rm sys} \otimes
\ket{\varphi}\bra{\varphi}$ and $\varrho_{\rm pr}(0) =
\ket{\varphi}\bra{\varphi}$ for some fixed vector $\ket{\varphi}
\in {\cal H}_{\rm pr}$. We suppose the projective measurements are
ideal and can be realized with a proper energy of the measuring
device~\cite{navascues-2014}. Rank-1 projectors are usually
considered in the analysis of Zeno and anti-Zeno effects; however,
our case crucially differs from those because the measurements are
performed on the probe and not on the system itself. Though the
probe dynamics is effectively frozen if $\gamma \tau \rightarrow
0$, the system continues evolving. If $\gamma(T - \lfloor T/ \tau
\rfloor \tau) \ll 1$, then the action of map $\Phi_T$ can be
approximated as follows:
\begin{equation}
\label{N-times} \Phi_T [\varrho] \approx \left( {\cal U}_{\tau}
\circ {\cal I}_i \right)^{T / \tau} [\varrho] = K \varrho
K^{\dag},
\end{equation}

\noindent where
\begin{eqnarray}
\label{K-rank-1}
&& K = \big( G(\gamma\tau) \big)^{T/\tau } , \\
&& G(\gamma\tau) = \sum_{k=0}^{\infty} \sum_{j_1,\ldots,j_k}
\frac{(-i \gamma \tau)^k}{k!} A_{j_1} \cdots A_{j_k} \big\langle
B_{j_1} \cdots B_{j_k} \big\rangle,\nonumber\\
&& \big\langle B_{j_1} \cdots B_{j_k} \big\rangle = \bra{\varphi}
B_{j_1} \cdots B_{j_k} \ket{\varphi}. \label{notation-average}
\end{eqnarray}

\noindent If $T / \tau$ is an integer, then Eq.~\eqref{N-times}
becomes exact. The approximate form of the Kraus operator $K$ can
be rewritten in the following form:
\begin{equation}
\label{K-simplified} K = \exp \Bigg\{ -iT \left[ \frac{i}{\tau}
\ln G(\tau\gamma) \right] \Bigg\},
\end{equation}

\noindent which can be simplified under the circumstances
\begin{equation}
\label{limit} \tau\rightarrow 0, \quad \gamma^2\tau = \Omega =
{\rm const}
\end{equation}

\noindent referred to as a \emph{stroboscopic limit}.

%%%%%%%%%%%%%%%%%%%%%%%%%%%%%%%%%%%%%%%%%%%%%%%%%%%%%%%%%%%%%%%%%%%
%%%%%%%%%%%%%%%%%%%%%%%%%%%%%%%%%%%%%%%%%%%%%%%%%%%%%%%%%%%%%%%%%%%

The physical meaning of the stroboscopic limit~\eqref{limit} is
that $\gamma \tau \rightarrow 0$, i.e. the probe dynamics is
effectively frozen (the quantum Zeno effect) but the system
dynamics is not frozen. The mathematical machinery is as follows.
Changing the variables $\gamma$ and $\tau$ by $\Omega = \gamma^2
\tau$ and $\varkappa = \sqrt{\tau}$, one can expand the logarithm
in Eq.~\eqref{K-simplified} in the vicinity of $\varkappa=0$:
\begin{eqnarray}
&& \!\!\!\!\!\!\!\!\!\! \ln G(\sqrt{\Omega} \, \varkappa) =
-i \sqrt{\Omega} \varkappa \sum_j A_j \langle B_j \rangle \nonumber\\
&& \!\!\!\!\!\!\!\!\!\! \quad  - \frac{\Omega \varkappa^2}{2}
\sum_{jk} A_j A_k \Big( \langle B_j B_k \rangle - \langle B_j
\rangle \langle B_k \rangle \Big) + O(\varkappa^3). \qquad
\label{logarithm}
\end{eqnarray}

\noindent Substituting Eq.~\eqref{logarithm} in
Eq.~\eqref{K-simplified} yields
\begin{equation}
K = e^{-i H_{\rm eff} T},
\end{equation}

\noindent where the effective Hamiltonian $H_{\rm eff}$ reads
\begin{eqnarray}
\label{H-effective-rank-1}
&& H_{\rm eff} = H_{1} - i H_{2} + O\left(\sqrt{\tau}\right), \\
&& H_{1} = \gamma \sum_j A_j \langle B_{j} \rangle, \\
&& H_{2} = \frac{\Omega}{2} \sum_{jk} A_j A_k \left(\langle B_{j}
B_{k} \rangle - \langle B_{j} \rangle \langle B_{k} \rangle
\right).
\end{eqnarray}

\noindent Both $H_{1}$ and $H_{2}$ are Hermitian and, besides,
$H_2 \geqslant 0$. In fact, it is not hard to see that $H_{2} =
\frac{\tau}{2} D_i^{\dag} D_i$, where $D_i = (I - C_i) H I_{\rm
sys} \otimes \ket{\varphi}$. The fact $H_2 \geqslant 0$ implies
the trace-decreasing property of the map \eqref{N-times}. Note
that $H_2 \neq 0$ if and only if $(I - C_i) H C_i \neq 0$, i.e.
the original Hamiltonian causes transitions between the sectors
${\rm supp}P_i$ and ${\rm ker}P_i$. The covariance matrix $M_{jk}
= \langle B_{j} B_{k} \rangle - \langle B_{j} \rangle \langle
B_{k} \rangle$ quantifies the intensity of such transitions.

The normalized system state $\varrho_{\rm sys}(T)$ satisfies the
non-linear equation
\begin{equation}
\label{non-linear-rank-1} \frac{\partial \varrho_{\rm
sys}}{\partial T} = -i \left[ H_1, \varrho_{\rm sys} \right] -
\left\{ H_2, \varrho_{\rm sys} \right\} + 2 {\rm tr} \left( H_2
\varrho_{\rm sys} \right) \varrho_{\rm sys},
\end{equation}

\noindent where $\{\cdot,\cdot\}$ denotes the anticommutator. The
purity parameter ${\rm tr}[\varrho_{\rm sys}^2]$ evolves
non-monotonically in general because the sign of derivative
\begin{equation}
\frac{\partial}{\partial T} {\rm tr}[\varrho_{\rm sys}^2] = 4
\left( {\rm tr}[\varrho_{\rm sys}^2] {\rm tr}[H_2 \varrho_{\rm
sys}] - {\rm tr}[H_2 \varrho_{\rm sys}^2] \right)
\end{equation}

\noindent depends on $H_2$ and $\varrho_{\rm sys}$.

Suppose the initial state $\varrho_{\rm sys}(0)$ is pure. Then the
approximate dynamics \eqref{non-linear-rank-1} preserves purity of
the initial state and the evolution of system state vector
$\ket{\psi}$ satisfies
\begin{equation}
\label{non-linear-psi-rank-1} i \frac{\partial
\ket{\psi}}{\partial T} = (H_1 - i H_2) \ket{\psi} + i \bra{\psi}
H_2 \ket{\psi} \, \ket{\psi}.
\end{equation}

\noindent Analogous equations are used in stochastic
interpretation of open quantum system
dynamics~\cite{breuer-petruccione-2002}. If we recall the exact
dynamics according to Eq.~\eqref{system-exact-evolution}, then at
each time $t=n\tau$, $n\in\mathbb{N}$, the exact density operator
$\varrho_{\rm sys}(n\tau)$ is also pure as a result of the
measurement performed. In between, the exact density operator
$\varrho_{\rm sys}(t)$ can become mixed but the less the product
$\gamma \tau$ the greater is the purity ${\rm tr}[\varrho_{\rm
sys}^2(t)]$. In the stroboscopic limit~\eqref{limit}, the exact
dynamics \eqref{system-exact-evolution} reduces to
Eqs.~\eqref{non-linear-rank-1}--\eqref{non-linear-psi-rank-1}.

Let us illustrate the developed theory for a physical system of
two coupled qubits (${\rm dim}{\cal H}_{\rm sys} = {\rm dim}{\cal
H}_{\rm pr} = 2$), one of which is being frequently measured with
a finite rate $\tau^{-1}$.

%%%%%%%%%%%%%%%%%%%%%%%%%%%%%%%%%%%%%%%%%%%%%%%%%%%%%%%%%%%%%%%%%%%
\begin{figure}
\includegraphics[width=8.5cm]{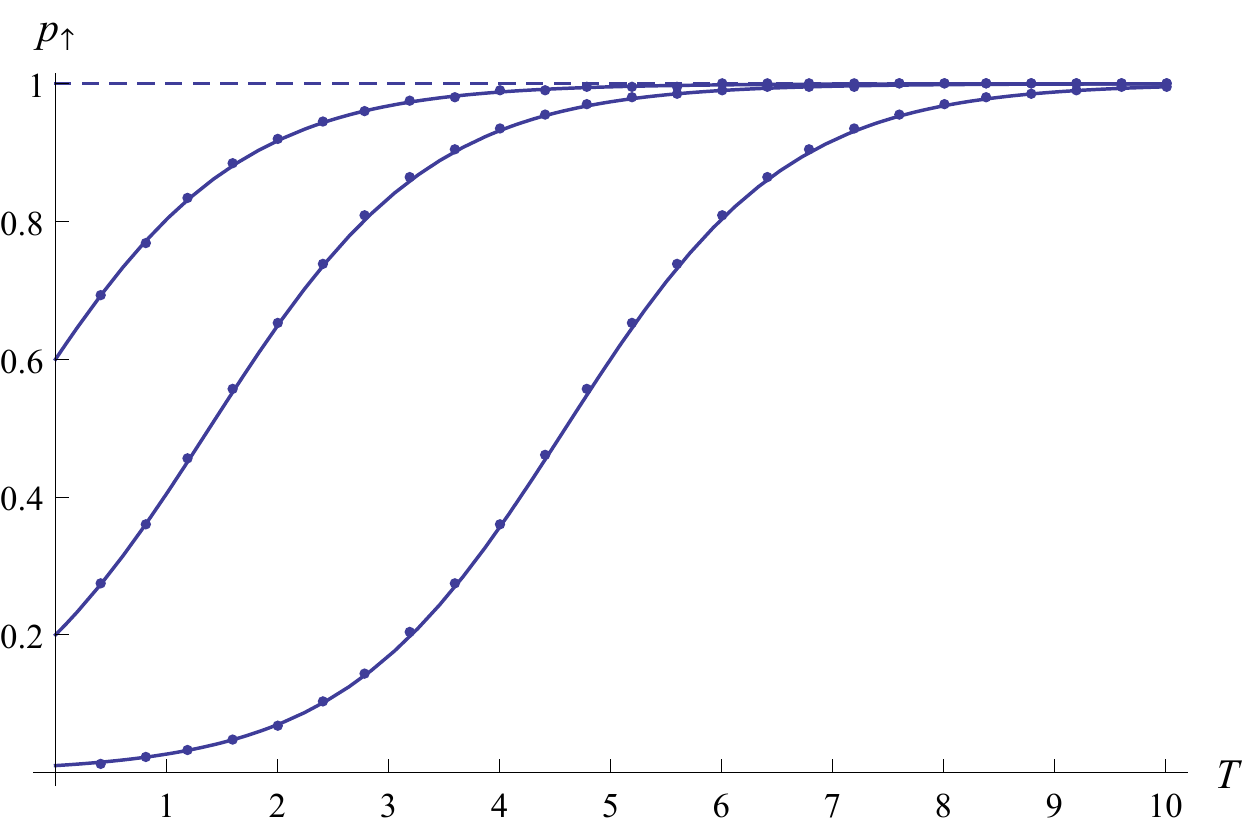}
\caption{\label{figure3} Comparison of exact (dots) and analytical
(solid line) expressions \eqref{probability-example} for
probability of finding the qubit system in the excited state at
time $T$ as a result of the stroboscopic
evolution~\eqref{system-exact-evolution} and its description via
the effective Hamiltonian~\eqref{H-effective-rank-1} with
$\gamma=5$, $\tau = 0.04$, $\Omega = 1$, and the initial state of
the system is $\ket{\psi(0)} = \alpha \ket{\uparrow} + \beta
\ket{\downarrow}$, where $|\alpha|^2 = 0.01$, $|\beta|^2 = 0.99$
(bottom line), $|\alpha|^2 = 0.2$, $|\beta|^2 = 0.8$ (middle
line), and $|\alpha|^2 = 0.6$, $|\beta|^2 = 0.4$ (top line).
Dashed line corresponds to the evolution with initial state
$\ket{\psi(0)} = \ket{\uparrow}$.}
\end{figure}
%%%%%%%%%%%%%%%%%%%%%%%%%%%%%%%%%%%%%%%%%%%%%%%%%%%%%%%%%%%%%%%%%%%

%%%%%%%%%%%%%%%%%%%%%%%%%%%%%%%%%%%%%%%%%%%%%%%%%%%%%%%%%%%%%%%%%%%
\begin{figure*}
\includegraphics[width=18cm]{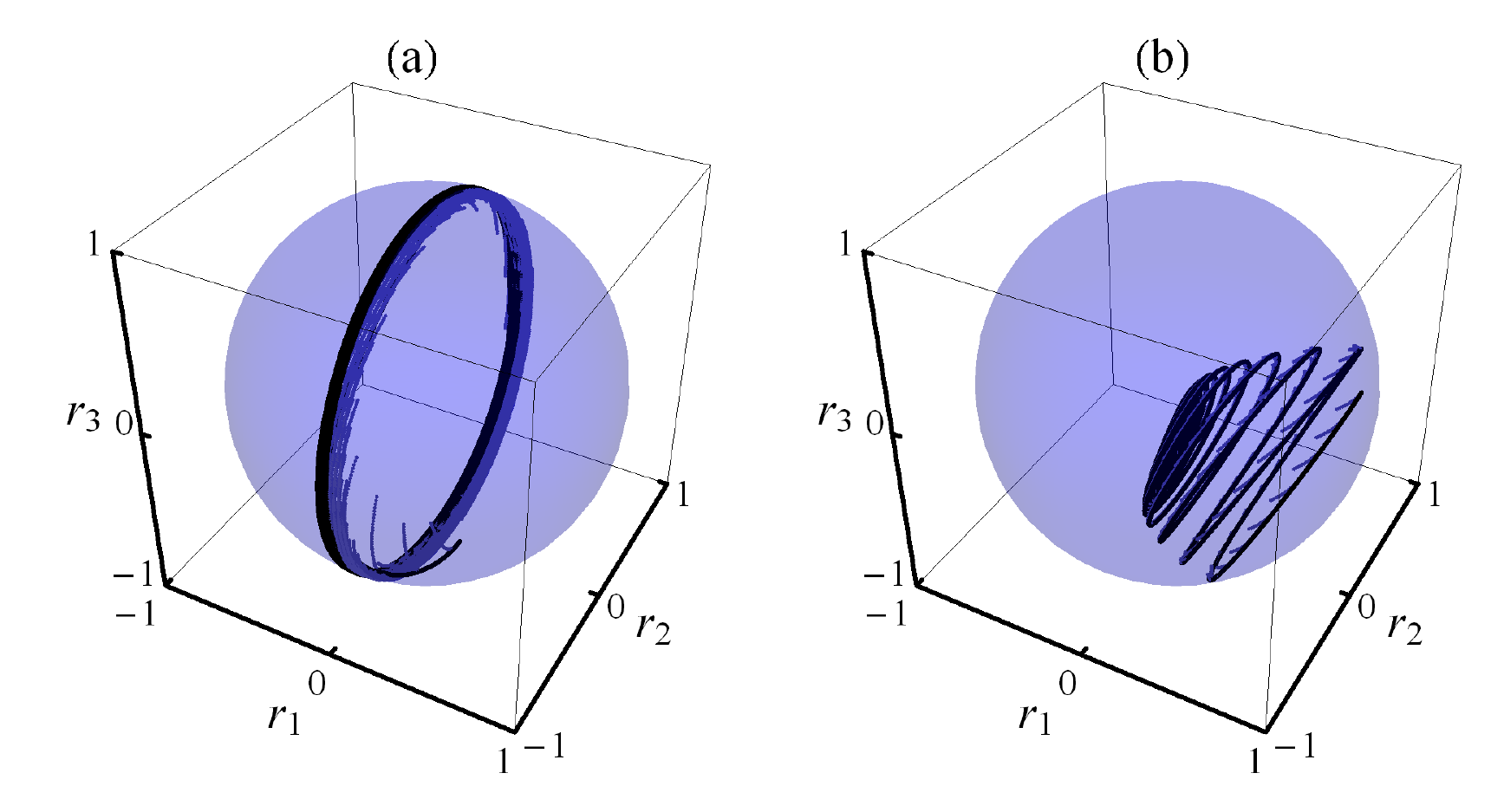}
\caption{\label{figure4} Qubit evolution $\varrho_{\rm sys}(t) =
\frac{1}{2}(I + \sum_{i=1}^3 r_i(t) \sigma_i)$ as a result of
system-probe nonlinear stroboscopic dynamics. Discontinuous line
depicts the exact evolution and emphasizes the stroboscopic
character of measurements [formulas \eqref{instrument-exact} and
\eqref{system-exact-evolution}]. Continuous black line is the
analytical approximate solution. (a) Limit cycle in
Example~\ref{example-2}, (b) dynamics inside the Bloch ball in
Example~\ref{example-3}.}
\end{figure*}
%%%%%%%%%%%%%%%%%%%%%%%%%%%%%%%%%%%%%%%%%%%%%%%%%%%%%%%%%%%%%%%%%%%

\begin{example}
Let two qubits interact with Hamiltonian $H=\gamma \text{\sc SWAP}
= \frac{\gamma}{2} \sum_{j=0}^{3} \sigma_j \otimes \sigma_j$,
where $\sigma_0 = I$ and $(\sigma_1,\sigma_2,\sigma_3)$ is the
conventional set of Pauli operators (Heisenberg Hamiltonian for
spins). By $U_{\tau}=e^{-i H \tau}$ denote the evolution operator
for a period $\tau$. Suppose $P_i = \varrho_{\rm pr}(0) =
\ket{\uparrow}\bra{\uparrow}$, where $\sigma_3\ket{\uparrow} =
\ket{\uparrow}$ and $\sigma_3\ket{\downarrow} = -
\ket{\downarrow}$, i.e. one stroboscopically measures the
particular spin projection of the probe qubit.  Then
\begin{equation}
U_{\tau}\ket{\psi}\otimes\ket{\uparrow} = \cos\gamma\tau
\ket{\psi} \otimes \ket{\uparrow} - i \sin\gamma\tau
\ket{\uparrow}\otimes\ket{\psi}
\end{equation}

\noindent and the Kraus operator for a period $\tau$ reads
$K_{\tau} = (I_{\rm sys}\otimes P_i) U_{\tau} = e^{-i\gamma\tau}
\ket{\uparrow}\bra{\uparrow} + \cos\gamma\tau
\ket{\downarrow}\bra{\downarrow}$. Since
$(\cos\gamma\tau)^{T/\tau} \rightarrow \exp\left(- \frac{1}{2}
\Omega T \right)$ in the stroboscopic limit \eqref{limit},
$K=K_{\tau}^{T/\tau} = \exp(-iH_{\rm eff}T)$, where the effective
system Hamiltonian reads
\begin{equation}
H_{\rm eff} = \gamma \ket{\uparrow}\bra{\uparrow} - i
\frac{\Omega}{2} \ket{\downarrow}\bra{\downarrow}.
\end{equation}

\noindent Formula \eqref{H-effective-rank-1} gives the same
result. Suppose the initial system state is $\ket{\psi(0)} =
\alpha \ket{\uparrow} + \beta \ket{\downarrow}$, $|\alpha|^2 +
|\beta|^2 = 1$, then at time $T=N\tau$ the probability to find the
system in the state $\ket{\uparrow}$ equals
\begin{equation}
\label{probability-example} p_{\uparrow} = \frac{ |\alpha|^2 }{
|\alpha|^2 + (\cos\gamma\tau)^{2N} |\beta|^2 } \approx \frac{
|\alpha|^2 }{ |\alpha|^2 + e^{-\Omega T} |\beta|^2 },
\end{equation}

\noindent where the left hand side of \eqref{probability-example}
is the exact expression and the right hand side of
\eqref{probability-example} is obtained via the effective
Hamiltonian. Comparison of exact numerical and approximate
analytical results is presented in Fig.~\ref{figure3}.

\end{example}

The non-linear effects become relevant at time $\sim \Omega^{-1}$
with the increase in the probability of error $p_{\rm err} = 1 -
p_{\Phi}$, which happens when the measurement outcome $i_k \ne i$.
In the example above, $p_{\rm err} = |\beta|^2(1-e^{-\Omega T})$.
Therefore, there is a trade-off between non-linearity and the
probability of its physical observation.

%----------------------------------------------------------------------

\subsection{\label{section-rank-r} Analytical solution for rank-$r$ projectors}

Let us now consider the case when $P_i$ in formula \eqref{C-P} is
a rank-$r$ projector. Physically it corresponds to a measurement
of an incomplete set of variables. For instance, measurement of
the square of the spin angular momentum, ${\bf s}^2$, and
observation of the outcome $s(s+1)$ does not specify the spin
projection $m =s,s-1,\ldots,-s$, i.e. $r = 2s+1$. Due to
degeneracy of $P_i$, the probe is not frozen (in contrast to
$r=1$), which is known as a Zeno subspace
effect~\cite{facchi-2002,li-2013}. The probe dynamics is
restricted to the subspace ${\rm supp}P_i$ and the subspace ${\rm
ker}P_i$ is forbidden (see Fig.~\ref{figure2}). Thus, the
system-probe dynamics takes place in the space ${\cal H}_{\rm sys}
\otimes {\cal H}_{r}$ and can be additionally simplified if the
condition \eqref{limit} is fulfilled.

Introduce the operators
\begin{equation}
G_{j} = P_i  B_j P_i, \qquad G_{jk} = P_i B_j B_k P_i,
\end{equation}

\noindent which generalize Eq.~\eqref{notation-average}.
Restricting ourselves to the subspace ${\rm supp} C_i$, we see
that the operator $C_i$ acts as the identity operator on all
vectors $\ket{\xi} \in {\rm supp}C_i$. In the domain ${\rm
supp}C_i$ we have
\begin{eqnarray}
&& \ln \left[ C_i - i\gamma\tau \sum_j G_j - \frac{\gamma^2 \tau^2}{2} \sum_{jk} G_{jk} + O\left((\gamma\tau)^3\right) \right] \nonumber\\
&& = - i\gamma\tau \sum_j G_j - \frac{\gamma^2 \tau^2}{2}
\sum_{jk} \left( G_{jk} - G_j G_k \right) +
O\left((\gamma\tau)^3\right). \nonumber
\end{eqnarray}

\noindent Continuing the same line of reasoning as in
Sec.~\ref{section-rank-1}, we obtain in the stroboscopic limit
\eqref{limit} that the system-probe density operator $\varrho$
evolves in the subspace ${\rm supp} C_i$ according to the
following equation:
\begin{equation}
\label{Phi-approximate-rank-r} \varrho(T) = \Phi_T [\varrho_{\rm
sys}(0) \otimes \varrho_{\rm pr}(0)] \approx e^{-i H_{\rm eff} T}
\varrho_{\rm sys}(0) \otimes \varrho_{\rm pr}(0) e^{i H_{\rm
eff}^{\dag} T},
\end{equation}

\noindent where the effective Hamiltonian $H_{\rm eff}$ reads
\begin{eqnarray}
\label{H-effective-rank-r}
&& H_{\rm eff} = H_{1} - i H_{2} + O\left(\sqrt{\tau}\right), \\
&& H_{1} = \gamma \sum_{j} A_{j}\otimes G_{j},\\
&& H_{2} = \frac{\Omega}{2} \sum_{jk} A_j A_k \otimes \left(
G_{jk} - G_{j}G_{k} \right).
\end{eqnarray}

By the same arguments, the term $H_2 = \frac{\tau}{2} {\rm
tr}_{\rm pr} [ C_i H (I-C_i) H C_i ] \geqslant 0$ is responsible
for trace decreasing.

Despite the fact that the system-probe dynamics is governed by the
effective Hamiltonian \eqref{H-effective-rank-r}, the dynamics of
system state $\varrho_{\rm sys}$ is not described by the effective
Hamiltonian in general and must be obtained via tracing out the
probe. Comparison of the exact system evolution [given by
Eqs.~\eqref{instrument-exact-arbitrary} and
\eqref{system-exact-evolution}] and the approximate analytical
system evolution [given by Eqs.~\eqref{Phi-approximate-rank-r} and
\eqref{system-exact-evolution}] is presented in the following
examples.

\begin{example}
\label{example-2} Consider three qubits (labelled $a$, $b$, $c$),
where qubit $a$ is a system and two qubits $b$ and $c$ represent a
probe. Let $P_i = \ket{S=0,M=0} \bra{S=0,M=0} + \ket{S=1,M=0}
\bra{S=1,M=0}$ be a projector onto subspace with zero total spin
projection of the probe. In conventional notation $P_i =
\ket{\uparrow_b \downarrow_c}\bra{\uparrow_b \downarrow_c} +
\ket{\downarrow_b \uparrow_c}\bra{\downarrow_b \uparrow_c}$, ${\rm
rank}P_i = 2$. Suppose the Heisenberg interaction Hamiltonian of
three qubits in local external fields, $H = \gamma
(\overrightarrow{\sigma}^a \overrightarrow{\sigma}^b +
\overrightarrow{\sigma}^b \overrightarrow{\sigma}^c +
\overrightarrow{\sigma}^c \overrightarrow{\sigma}^{a} +
\sigma_x^{a} + \sigma_y^b + \sigma_z^c)$. Then stroboscopic
measurements with $\tau=0.04$ and $\gamma = 5$ satisfy $\gamma
\tau \ll 1$, and the stroboscopic theory can be applied.
Comparison of the exact and approximate dynamics of qubit $a$ is
depicted in Fig.~\ref{figure4}(a). Dynamics represents a limit
cycle, which manifests the non-linear character of evolution.
\end{example}

\begin{example}
\label{example-3} In the above example of three qubits (labelled
$a$, $b$, $c$), change the projector $P_i = \ket{S=1,M=1}
\bra{S=1,M=1} + \ket{S=1,M=-1} \bra{S=1,M=-1} = \ket{\uparrow_b
\uparrow_c}\bra{\uparrow_b \uparrow_c} + \ket{\downarrow_b
\downarrow_c}\bra{\downarrow_b \downarrow_c}$ and consider
Heisenberg interaction Hamiltonian of three qubits in a global
external field, $H = \gamma (\overrightarrow{\sigma}^a
\overrightarrow{\sigma}^b + \overrightarrow{\sigma}^b
\overrightarrow{\sigma}^c + \overrightarrow{\sigma}^c
\overrightarrow{\sigma}^{a} + \sigma_z^{a} + \sigma_z^b +
\sigma_z^c)$. Stroboscopic measurements with $\tau=0.02$ and
$\gamma = 2\sqrt{2}$ satisfy $\gamma \tau \ll 1$, and one can use
the approximate formula of the stroboscopic limit. Comparison of
the exact and approximate dynamics of qubit $a$ inside the Bloch
ball is depicted in Fig.~\ref{figure4}(b). The decrease of the
system state purity can be attributed to the entanglement between
the system and the probe. In fact, the
map~\eqref{Phi-approximate-rank-r} is bipartite with respect to
the system and the probe; entanglement preserving and entanglement
annihilating properties of bipartite maps are
characterized~\cite{filippov-2013}.
\end{example}

The above examples show that the analytical approximate results
are in perfect agreement with the exact numerical ones.

%----------------------------------------------------------------------

\section{\label{section-non-selective} Stroboscopic limit for non-selective measurements}

For the sake of generality we do not single out the system and the
probe in the beginning of this section, so we deal with a general
Hilbert space ${\cal H}_{\rm sys+pr}$. Let us consider a
non-selective projective measurement described by the
trace-preserving completely positive map (measurement channel)
\begin{equation}
\label{lambda-map} \Lambda [\varrho] = \sum_{i=1}^{m} C_i \varrho
C_i,
\end{equation}

\noindent where each operator $C_i = C_i^2$ is a projector and $m$
is a number of Kraus operators, $\sum_{i=1}^m C_i = I$. There is
no restriction on the dimension of projectors, i.e. ${\rm dim} \,
{\rm supp} C_i$ is arbitrary, with $\sum_{i=1}^m {\rm dim} \, {\rm
supp} C_i = {\rm dim} {\cal H}_{\rm sys + pr}$. Note that
\begin{equation}
\label{Proj.} \Lambda \circ \Lambda = \Lambda
\end{equation}

\noindent since $C_i C_j = \delta_{ij} C_j$.

%%%%%%%%%%%%%%%%%%%%%%%%%%%%%%%%%%%%%%%%%%%%%%%%%%%%%%%%%%%%%%%%%%%
\begin{figure}
\includegraphics[width=8.5cm]{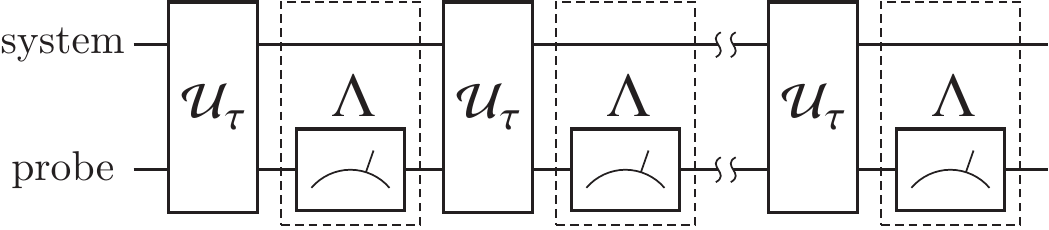}
\caption{\label{figure5} Stroboscopic non-selective measurements.
Though the unitary evolution is interrupted by measurements
\eqref{lambda-map}, the dynamics is not frozen in ${\rm
supp}\Lambda$.}
\end{figure}
%%%%%%%%%%%%%%%%%%%%%%%%%%%%%%%%%%%%%%%%%%%%%%%%%%%%%%%%%%%%%%%%%%%

Suppose non-selective measurements \eqref{lambda-map} are
performed successively after equal time intervals of duration
$\tau$, with the intermediate unitary evolution being described by
Eq.~\eqref{unitary}, see Fig.~\ref{figure5}. Such a stroboscopic
dynamics with non-selective measurements defines the dynamical map
\begin{equation}
\label{Lambda-exact-arbitrary} \Phi_{T} = {\cal U}_{T - N \tau}
\circ \Lambda \circ \ldots \circ \Lambda \circ {\cal U}_{\tau}
\circ \Lambda \circ {\cal U}_{\tau},
\end{equation}

\noindent which describes the exact evolution. It is natural to
assume that the initial state $\varrho(0)$ is an eigenstate of
$\Lambda$, i.e. $\Lambda[\varrho(0)] = \varrho(0)$. If this is not
the case, one can set $t=0$ at the moment of the first
measurement.

At times $T=n\tau$, $n\in\mathbb{N}$, the transformation
$\Phi_{T}$ maps any operator into ${\rm supp}\Lambda$. For
instance, the density matrix at such time moments has the form
\begin{equation}
\label{block-diagonal} \varrho(T) = \left(
\begin{array}{cccc}
\blockmatrix{0.45in}{0.3in}{$\rho^{(1)}(T)$} & \multicolumn{3}{|c}{}\\
\cline{1-2} \multicolumn{1}{c|}{} &
\blockmatrix{0.45in}{0.3in}{$\rho^{(2)}(T)$} &\multicolumn{2}{|c}
{\raisebox{1.5ex}[0pt]{\parbox{12pt}{\Huge 0}}}\\
\cline{2-2} \multicolumn{2}{c}{}& \ddots &   \\
\cline{4-4} \multicolumn{3}{c|}
{\raisebox{-1.0ex}[0pt]{\parbox{12pt}{\Huge 0}}} &
\blockmatrix{0.45in}{0.3in}{$\rho^{(m)}(T)$}
\end{array}
 \right)
\end{equation}

\noindent in the basis of eigenvectors of $C_i$, where
$\rho^{(i)}(T)$ is a restriction of the operator
\begin{equation}
\label{rho-i} \varrho^{(i)}(T) = C_i \varrho(T) C_i
\end{equation}

\noindent to the subspace ${\rm supp}C_i$. Eq.~\eqref{rho-i} would
represent a conditional density operator if the outcome $i$ was
observed.  In other words, $\rho^{(i)}(T)$ is a non-zero minor of
the matrix representation of $\varrho^{(i)}(T)$ in the basis of
eigenvectors of $C_i$.

In what follows we derive analytical equations for the approximate
dynamics of $\varrho(T)$ in ${\rm supp}\Lambda$ in the
stroboscopic limit.

For the sake of generality we do not impose any restrictions on
the Hamiltonian $H$ except extracting its characteristic strength
explicitly, i.e. $H = \gamma h$, where $h$ is dimensionless and
its operator norm $\| h \|_{\infty} \leqslant 1$. Then the unitary
map ${\cal U}_t = \exp(\gamma {\cal L} t)$, where the generator
${\cal L}$ reads
\begin{equation}
\label{L-exact} {\cal L} [\varrho] = -i [h,\varrho].
\end{equation}

Approximate formula for the dynamical map is
\begin{equation}
\Phi_T = \Lambda \circ \exp[\gamma\tau{\cal L}] \circ \Lambda
\circ \ldots \circ \Lambda \circ \exp[\gamma\tau{\cal L}] \circ
\Lambda,
\end{equation}

\noindent where we have taken into account that
$\Lambda[\varrho(0)] = \varrho(0)$. Using the property
(\ref{Proj.}), we get
\begin{eqnarray}
\Phi_T &=& \left[ \Lambda \circ \exp \left( \gamma \tau {\cal L}
\right) \circ \Lambda \right]^{T / \tau}
= \left[ \sum_{k=0}^{\infty} \frac{(\tau\gamma)^k}{k!} \Lambda \circ {\cal L}^k \circ \Lambda \right]^{T / \tau} \nonumber\\
&=& \exp \left\{ T \left[ \frac{1}{\tau} \ln \left(
\sum\limits_{k=0}^{\infty} \frac{(\tau\gamma)^k}{k!} \Lambda \circ
{\cal L}^k \circ \Lambda \right) \right] \right\}.
\end{eqnarray}

\noindent Some algebra in the stroboscopic limit $\gamma^2\tau =
\Omega$, $\tau \rightarrow 0$ yields
\begin{equation}
\label{Phi-T-non-selective} \Phi_T = \exp \left( {\cal L}_{\rm
eff} T \right),
\end{equation}

\noindent where the effective dynamical semigroup
generator~\cite{gks-1976,lindblad-1976} reads
\begin{equation}
\label{effective-generator} {\cal L}_{\rm eff} =
\gamma\Lambda\circ{\cal L} \circ \Lambda + \frac{\Omega}{2} \Big(
\Lambda \circ {\cal L}^2 \circ \Lambda - \Lambda \circ {\cal L}
\circ \Lambda \circ{\cal L} \circ \Lambda \Big).
\end{equation}

The fact that the measurement-induced dynamics reduces to a
dynamical semigroup in the stroboscopic limit is in agreement with
the earlier studies, where master equations were derived for
calculation of the modified decay
rates~\cite{kofman-2000,kofman-2001}.

To get the particular form of the generator
\eqref{effective-generator} for the system-probe Hamiltonian $H =
\gamma h$, we introduce auxiliary operators
\begin{equation}
h_{ij} = C_i h C_j,
\end{equation}

\noindent whose physical meaning is the transition between
measurement-invariant subspaces ${\rm supp} C_j$ and ${\rm supp}
C_i$. Note that $h_{ij}$ denotes an operator not a matrix element.
We substitute \eqref{L-exact} into each term of
\eqref{effective-generator} and expand
\begin{eqnarray}
&& \Lambda \circ{\cal L} \circ \Lambda[\varrho] = -i \sum_{i}
\left[
h_{ii}, \varrho \right], \\
&& \Lambda \circ {\cal L} \circ \Lambda \circ {\cal L} \circ
\Lambda[\varrho] = -\sum_{i} \left\{ h_{ii}^2, \varrho \right\} -
2 h_{ii} \varrho h_{ii}, \\
&& \Lambda \circ {\cal L}^2 \circ \Lambda [\varrho] = -\sum_{ij}
\{ h_{ji}^{\dag} h_{ji}, \varrho \} - 2 h_{ji} \varrho
h_{ji}^{\dag}
\end{eqnarray}

\noindent to get the final expression for the effective dynamical
semigroup generator in the Lindblad form:
\begin{equation}
{\cal L}_{\rm eff} [\varrho] = -i \gamma \sum_{i} \left[ h_{ii},
\varrho \right] - \frac{\Omega}{2} \sum_{i\ne j}\Big( \left\{
h_{ji}^\dag h_{ji}, \varrho \right\} - 2 h_{ji} \varrho
h_{ji}^\dag \Big).
\end{equation}

\noindent One can see that the non-transition operators $h_{ii}$
are responsible for the unitary evolution, whereas the transition
operators $h_{ij}$, $i \neq j$ are exactly the Lindblad operators
responsible for the dissipation and decoherence. Because
$h_{ji}^{\dag} = h_{ij}$, the global density operator $\varrho$
evolution
\begin{eqnarray}
\label{Lindblad-equation} \frac{\partial \varrho}{\partial T} =
{\cal L}_{\rm eff} \varrho
\end{eqnarray}

\noindent preserves the block-diagonal structure
[\eqref{block-diagonal}] of the density operator, i.e. $\varrho(T)
= \sum_i \varrho^{(i)}(T)$. Therefore,
Eq.~\eqref{Lindblad-equation} reduces to
\begin{eqnarray}
\label{evolution-of-blocks} \frac{\partial}{\partial T} \sum_{i}
\varrho^{(i)} &=& -i \gamma \sum_{i} \left[
h_{ii},\varrho^{(i)} \right] \nonumber\\
&& - \frac{\Omega}{2} \sum_{i \ne j} \Big( \left\{ h_{ij} h_{ji},
\varrho^{(i)} \right\} - 2 h_{ij} \varrho^{(j)} h_{ji} \Big).
\qquad
\end{eqnarray}

\noindent Taking into account that $\sum_{i\ne j} C_j = I - C_i$,
we simplify
\begin{eqnarray}
\label{summation-simplification} \sum_{i\ne j} \left\{ h_{ij}
h_{ji}, \varrho^{(i)} \right\} &=& \left\{ C_i h (I - C_i) h C_i,
\varrho^{(i)} \right\}
\nonumber\\
&=& \left\{ (h^2)_{ii} - (h_{ii})^2, \varrho^{(i)} \right\}
\end{eqnarray}

\noindent and obtain the Hamiltonian dispersion in each block. The
operator $(h^2)_{ii} - (h_{ii})^2$ does not vanish only if
transition terms $h_{ij}$ are non-zero. Physically, transition
terms cause short-period correlations between ${\rm supp}C_j$ and
${\rm supp}C_i$, which are then destroyed by a measurement.
Cancellation of those correlations leads to the decoherence in
diagonal blocks $\varrho^{(i)}$, and the term $(h^2)_{ii} -
(h_{ii})^2$ quantitatively describes such a decoherence.

Substituting \eqref{summation-simplification} in
\eqref{evolution-of-blocks}, we obtain
\begin{eqnarray}
\label{evolution-of-blocks-simplified} \frac{\partial}{\partial T}
\sum_{i} \varrho^{(i)} &=& -i \sum_{i} \Big(H^{\rm
eff}_{i}\varrho^{(i)}-\varrho^{(i)} \left(H_{i}^{\rm
eff}\right)^\dag\Big) \nonumber \\
&& + \Omega \sum_{i\ne j} h_{ij} \varrho^{(j)} h_{ji},
\end{eqnarray}

\noindent where the effective non-Hermitian Hamiltonian reads
\begin{equation}
H^{\rm eff}_{i}=\gamma
h_{ii}-\frac{i\Omega}{2}\Big((h^2)_{ii}-(h_{ii})^2\Big).
\end{equation}

\noindent The first term in the right-hand side of
Eq.~\eqref{evolution-of-blocks-simplified} can be interpreted as
the sum of individual selective evolutions of individual blocks
$\varrho^{(i)}$ in accordance with the results of
Sec.~\ref{section-problem}. The second term in the right-hand side
of Eq.~\eqref{evolution-of-blocks-simplified} describes the mutual
influence of the blocks.

Note that the maximally mixed state $\varrho = I / {\rm dim}{\cal
H}_{\rm sys+pr}$ is a fixed point of the dynamical map
\eqref{Phi-T-non-selective}.

Suppose that all projectors $C_i$ are one-dimensional, i.e. $C_i =
\ket{i}\bra{i}$, then the density operator \eqref{block-diagonal}
is diagonal, with diagonal elements being the probabilities
$p_i(T)$ to observe the outcome $i$. According to
Eq.~\eqref{evolution-of-blocks-simplified}, the evolution of these
probabilities has the form of the classical Pauli equation:
\begin{eqnarray}
&& \label{Pauli-eq} \frac{\partial p_i(T)}{\partial T} =
\sum_{j \ne i} \big( W_{j \rightarrow i} \, p_j(T) - W_{i \rightarrow j} \, p_i(T) \big), \\
&& W_{j \rightarrow i} = \Omega |\bra{i} h \ket{j}|^2, \\
&& \sum_{j \ne i} W_{i \rightarrow j} = \Omega \left( \bra{i} h^2
\ket{i} - \bra{i} h \ket{i}^2 \right).
\end{eqnarray}

\noindent Conceptually, Eq.~\eqref{Pauli-eq} shows that the
classical dynamics can be reproduced from the quantum dynamics in
the stroboscopic limit of non-selective rank-1 projective
measurements of the global system.

%%%%%%%%%%%%%%%%%%%%%%%%%%%%%%%%%%%%%%%%%%%%%%%%%%%%%%%%%%%%%%%%%%%
\begin{figure}
\includegraphics[width=6cm,height=6cm]{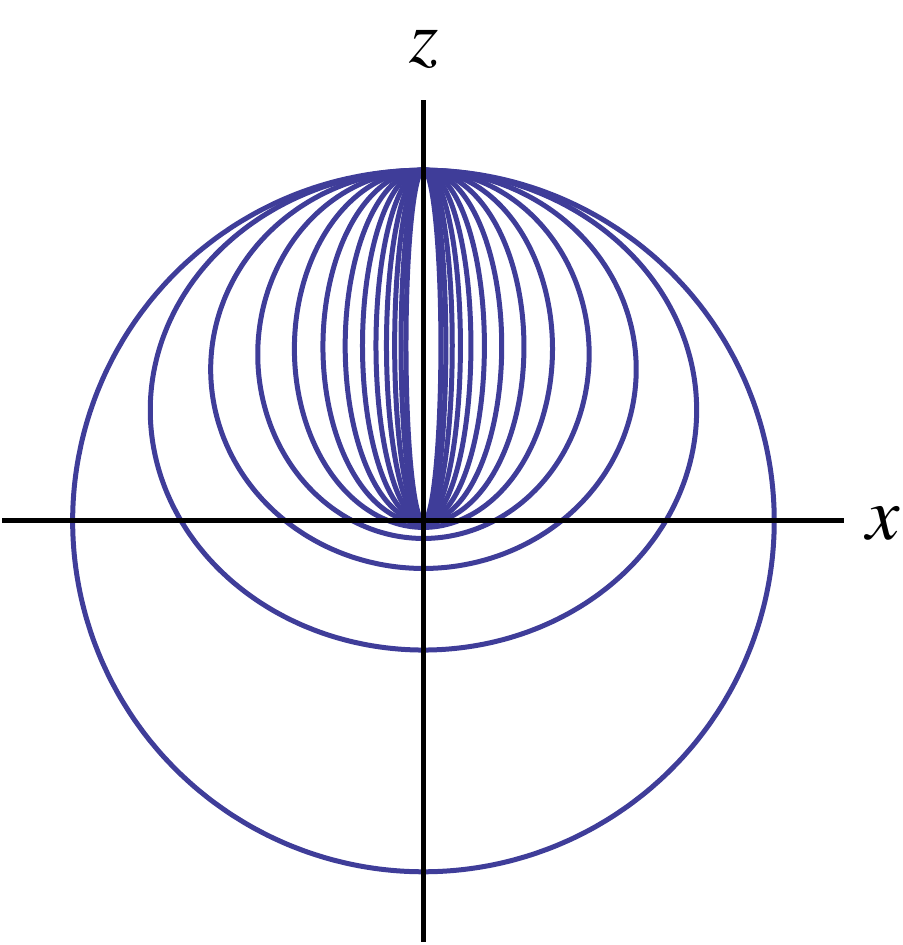}
\caption{\label{figure6} Evolution of the Bloch ball according to
Eq.~\eqref{non-selective-example-evolution},
Example~\ref{example-4}, which describes the qubit system
evolution when the coupled probe is stroboscopically measured.
Parameter $\Omega =0.1$, snapshot interval $\Delta T = 5$.}
\end{figure}
%%%%%%%%%%%%%%%%%%%%%%%%%%%%%%%%%%%%%%%%%%%%%%%%%%%%%%%%%%%%%%%%%%%

%----------------------------------------------------------------

\subsection{\label{section-non-selective-system} Dynamics of the system and the probe}

We now take into account the tensor product structure of the
Hilbert space ${\cal H}_{\rm sys + pr} = {\cal H}_{\rm sys}
\otimes {\cal H}_{\rm pr}$. Non-selective projective measurements
of the probe correspond to operators $C_i = I_{\rm sys} \otimes
P_i$, where $P_i = P_i^{\dag} = P_i^2$, $\sum_i P_i = I_{\rm pr}$,
and $\sum_i {\rm rank} P_i = {\rm dim} {\cal H}_{\rm pr}$, see
Fig.~\ref{figure5}.

Evolution of the system density matrix reads
\begin{equation}
\varrho_{\rm sys}(T) = \sum_i {\rm tr}_{\rm pr} \varrho^{(i)}(T)
\end{equation}

\noindent and cannot be reduced to the closed formula
$\frac{\partial}{\partial T} \varrho_{\rm sys} = {\cal L}_{\rm
sys} \varrho_{\rm sys}$ involving $\varrho_{\rm sys}$ only, as the
partial trace of the semigroup dynamics is not a semigroup
dynamics in general. The probe density matrix evolution exhibits
the same property.

If $P_i = \ket{i}_{\rm pr}\bra{i}$ for all $i$, then the probe
density operator is diagonal $\varrho_{\rm pr}(T) = \sum_i p_i(T)
\ket{i}_{\rm pr}\bra{i}$ and $p_i(T) = {\rm tr} \varrho^{(i)}(T)$.

\begin{example}
\label{example-4} Consider a two-qubit system, where the first
qubit plays the role of a system, and the second qubit is a probe.
Let the total Hamiltonian be $H = \gamma h = \gamma \text{\sc
SWAP} = \frac{\gamma}{2} \sum_{j=0}^3 \sigma_j \otimes \sigma_j$,
then $h^2 = I \otimes I$. Suppose non-selective projective
measurement of the probe qubit in the conventional basis
$\ket{\uparrow},\ket{\downarrow}$, then $C_1 = I \otimes
\ket{\uparrow}\bra{\uparrow}$ and $C_2 = I \otimes
\ket{\downarrow}\bra{\downarrow}$. A direct calculation yields
\begin{eqnarray}
&& \label{h11} h_{11} = \ket{\uparrow}\bra{\uparrow} \otimes
\ket{\uparrow}\bra{\uparrow}, \\
&& h_{12} = h_{21}^{\dag} = \ket{\downarrow}\bra{\uparrow} \otimes
\ket{\uparrow}\bra{\downarrow},\\
&& \label{h22} h_{22} = \ket{\downarrow}\bra{\downarrow} \otimes
\ket{\downarrow}\bra{\downarrow}.
\end{eqnarray}

\noindent Substituting Eqs.~\eqref{h11}--\eqref{h22} in
Eq.~\eqref{evolution-of-blocks-simplified}, we get the approximate
dynamics of the blocks $\varrho^{(1)}(T) = \rho^{(1)}(T) \otimes
\ket{\uparrow}\bra{\uparrow}$ and $\varrho^{(2)}(T) =
\rho^{(2)}(T) \otimes \ket{\downarrow}\bra{\downarrow}$
constituting the global system+probe density operator
$\varrho(T)$:
\begin{eqnarray}
\label{eq-rho-11} && \!\!\!\!\!\!\!\!\!\! \frac{\partial}{\partial T} \left(%
\begin{array}{cc}
  \rho^{(1)}_{\uparrow\uparrow} & \rho^{(1)}_{\uparrow\downarrow} \\
  \rho^{(1)}_{\downarrow\uparrow} & \rho^{(1)}_{\downarrow\downarrow} \\
\end{array}%
\right) = \left(%
\begin{array}{cc}
  0 & (-i\gamma - \Omega /2)\rho^{(1)}_{\uparrow\downarrow} \\
  (i\gamma - \Omega /2)\rho^{(1)}_{\downarrow\uparrow} & -\Omega \big( \rho^{(1)}_{\downarrow\downarrow} - \rho^{(2)}_{\uparrow\uparrow} \big) \\
\end{array}%
\right), \nonumber\\ \\
\label{eq-rho-22} && \!\!\!\!\!\!\!\!\!\! \frac{\partial}{\partial T} \left(%
\begin{array}{cc}
  \rho^{(2)}_{\uparrow\uparrow} & \rho^{(2)}_{\uparrow\downarrow} \\
  \rho^{(2)}_{\downarrow\uparrow} & \rho^{(2)}_{\downarrow\downarrow} \\
\end{array}%
\right) = \left(%
\begin{array}{cc}
  -\Omega \big( \rho^{(2)}_{\uparrow\uparrow} - \rho^{(1)}_{\downarrow\downarrow} \big) & (i\gamma - \Omega /2)\rho^{(1)}_{\uparrow\downarrow} \\
  (-i\gamma - \Omega /2)\rho^{(1)}_{\downarrow\uparrow} & 0 \\
\end{array}%
\right). \nonumber\\
\end{eqnarray}

\noindent Suppose the initial state of the probe is
$\ket{\uparrow}\bra{\uparrow}$ and the initial state of the system
is an arbitrary $2 \times 2$ density matrix $\varrho_{\rm
sys}(0)$, then $\rho^{(1)}(0) = \varrho_{\rm sys}(0)$ and
$\rho^{(2)}(0) = 0$. Solving the system of linear equations
\eqref{eq-rho-11}--\eqref{eq-rho-22} and summing $\varrho_{\rm
sys}(T) = \rho^{(1)}(T) + \rho^{(2)}(T)$, we get the system
density matrix evolution:
\begin{eqnarray}
\label{non-selective-example-evolution} && \!\!\!\!\!\!\!\!\!\! \varrho_{\rm sys}(T) \nonumber\\
&& \!\!\!\!\!\!\!\!\!\! = \left(%
\begin{array}{cc}
  \varrho_{\rm sys}^{\uparrow\uparrow}(0) \! + \! \frac{1}{2}\left( 1 \! - \! e^{-2 \Omega T}\right) \varrho_{\rm sys}^{\downarrow\downarrow}(0) & e^{(-i\gamma - \Omega / 2) T} \varrho_{\rm sys}^{\uparrow\downarrow}(0) \\
  e^{(i\gamma - \Omega / 2) T} \varrho_{\rm sys}^{\downarrow\uparrow}(0) & \frac{1}{2}\left( 1 \! + \! e^{-2 \Omega T}\right) \varrho_{\rm sys}^{\downarrow\downarrow}(0) \\
\end{array}%
\right). \nonumber\\
\end{eqnarray}

%%%%%%%%%%%%%%%%%%%%%%%%%%%%%%%%%%%%%%%%%%%%%%%%%%%%%%%%%%%%%%%%%%%
\begin{figure}
\includegraphics[width=8.5cm]{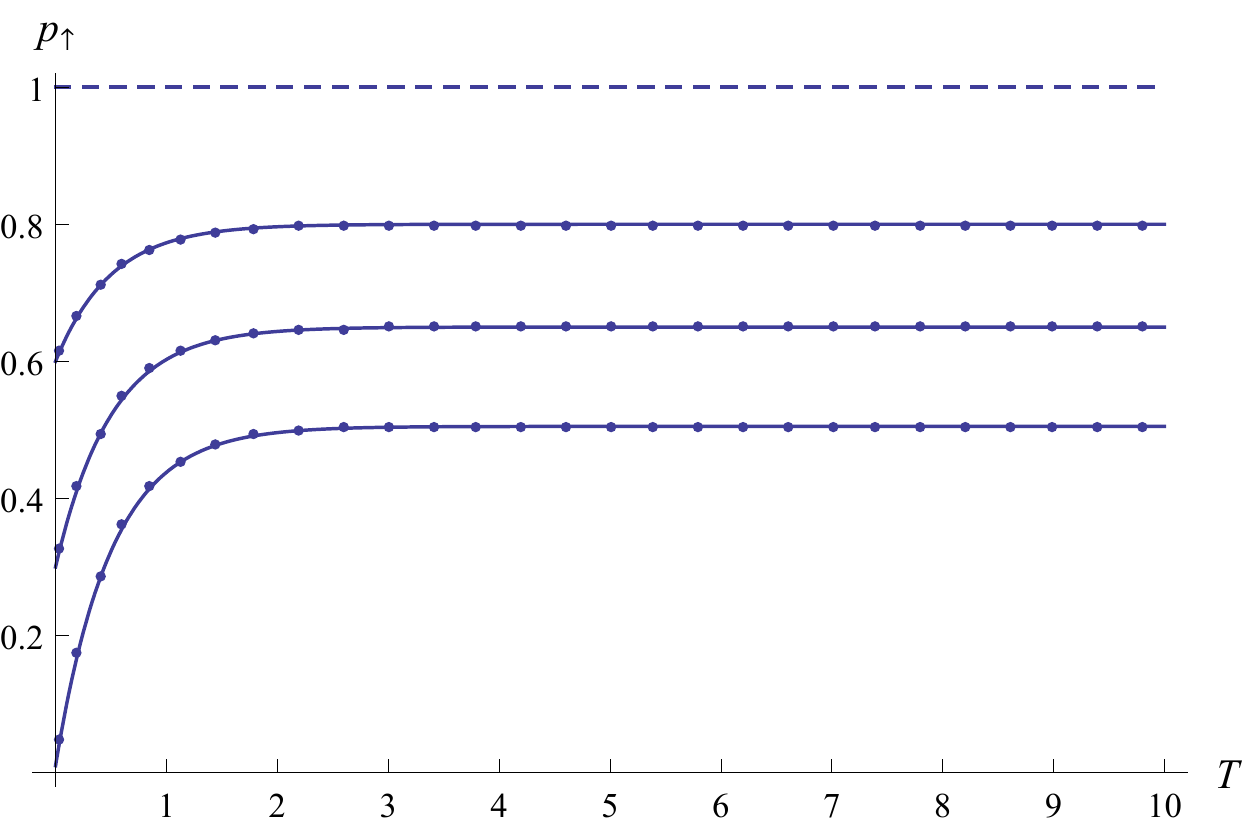}
\caption{\label{figure7} Comparison of exact (dots) and analytical
(solid line) probabilities of finding the qubit system in the
excited state at time $T$ as a result of the stroboscopic
evolution~\eqref{Lambda-exact-arbitrary} and its approximate
form~\eqref{non-selective-example-evolution} with $\gamma=5$,
$\tau = 0.04$, $\Omega = 1$, initial state of the system is
$\ket{\psi(0)} = \alpha \ket{\uparrow} + \beta \ket{\downarrow}$,
where $|\alpha|^2 = 0.01$, $|\beta|^2 = 0.99$ (bottom line),
$|\alpha|^2 = 0.3$, $|\beta|^2 = 0.7$ (middle line), and
$|\alpha|^2 = 0.6$, $|\beta|^2 = 0.4$ (top line). Dashed line
corresponds to the evolution with initial state $\ket{\psi(0)} =
\ket{\uparrow}$.}
\end{figure}
%%%%%%%%%%%%%%%%%%%%%%%%%%%%%%%%%%%%%%%%%%%%%%%%%%%%%%%%%%%%%%%%%%%

The transformation of the Bloch ball via
Eq.~\eqref{non-selective-example-evolution} is depicted in
Fig.~\ref{figure6}. One can readily see that the
evolution~\eqref{non-selective-example-evolution} does not have a
semigroup property, although the global system-probe dynamics
$\rho^{(1)}(T)\otimes\ket{\uparrow}\bra{\uparrow} +
\rho^{(2)}(T)\otimes\ket{\downarrow}\bra{\downarrow}$ has the
semigroup property.

The exact solution of the interrupted evolution
\eqref{Lambda-exact-arbitrary} is rather involved for the
considered example and no concise closed formula can be obtained.
The larger is the number of measurements $N = \lfloor T / \tau
\rfloor $, the more complicated is the calculation. We compare the
approximate analytical solution
\eqref{non-selective-example-evolution} with the exact numerical
solutions in Fig.~\ref{figure7}, which shows the agreement between
them.
\end{example}

%-------------------------------------------------------------------

\section{\label{section-conclusions} Conclusions}

We have considered the quantum system dynamics under frequent
successive measurements with a finite repetition rate $\tau^{-1}$
and derived approximate analytical evolution equations for the
timescale $T \sim (\gamma^2 \tau)^{-1}$, where $\gamma$ is the
characteristic strength of the Hamiltonian.

In the case of selective rank-1 projective measurements of the
probe, we have obtained the non-linear system evolution equation
for the most probable sequence of coincident outcomes. The pure
quantum states remain pure in such an evolution. For selective
rank-$r$ projective measurements of the probe, the system
evolution becomes more involved and may lead to the change of
purity, though the dynamics equation remains non-linear.

In the case of non-selective measurements, the stroboscopic limit
$\gamma^2 \tau = \Omega$, $\tau \rightarrow 0$ results in a
general Gorini-Kossakowski-Sudarshan-Lindblad equation for the
system-probe aggregate, though the reduced evolution of the system
may not exhibit the semigroup property. Finally, a classical
stochastic Pauli equation is obtained for non-selective projective
rank-1 measurements of the system-probe aggregate.

The obtained results can be treated as deviations from the Zeno
subspace dynamics, when the interval between measurements $\tau$
tends to zero and the number of measurements $N = T / \tau \sim
(\gamma \tau)^{-2}$ tends to infinity. The stroboscopic limit
provides analytical equations, which can be used in the analysis
of experiments with a high but finite repetition rate of repeated
measurements at timescale $T \sim (\gamma^2 \tau)^{-1}$. The
examples provided show the agreement between the exact numerical
dynamics and the approximate analytical one whenever $\gamma \tau
\ll 1$.

\begin{acknowledgements}
The authors thank Luigi Accardi for fruitful discussions. The
authors are grateful to the anonymous referee and Pavel Pyshkin
for valuable comments. The study in Sec.~\ref{section-problem} was
supported by Russian Science Foundation under project No.
16-11-00084 and performed in Moscow Institute of Physics and
Technology. The results of
section~\ref{section-non-selective-system} were obtained by S.N.
Filippov, supported by the Russian Foundation for Basic Research
under Project No. 16-37-60070 mol\_a\_dk, and performed at the
Institute of Physics and Technology of the Russian Academy of
Sciences.
\end{acknowledgements}

\end{document}